\documentclass[reprint,apl,aip,superscriptaddress]{revtex4-1}

\usepackage{float}
\usepackage{natbib}
\usepackage{amsmath}
\usepackage{graphicx}
\usepackage{times}
\usepackage{here}
\usepackage{graphics}
\usepackage{epsfig,graphicx}
\usepackage[version=3]{mhchem}

\begin{document}

\bibliographystyle{apsrev4-1}

\title{NO-assisted molecular-beam epitaxial growth of nitrogen substituted EuO}

\author{R. Wicks}
\affiliation{Department of Physics and Astronomy, University of British Columbia, Vancouver, British Columbia V6T 1Z1, Canada}

\author{S.~G.~Altendorf}
\altaffiliation[Present address: ]{Max Planck Institute for Chemical Physics of Solids, N\"othnitzer Str. 40, 01187 Dresden, Germany}
  \affiliation{II. Physikalisches Institut, Universit\"{a}t zu K\"{o}ln, Z\"{u}lpicher Str. 77, 50937 K\"{o}ln, Germany}

\author{C. Caspers}
\altaffiliation[Present address: ]{Peter Gr\"{u}nberg Institut (PGI-6), Forschungszentrum J\"{u}lich, 52425 J\"{u}lich, Germany}
    \affiliation{II. Physikalisches Institut, Universit\"{a}t zu K\"{o}ln, Z\"{u}lpicher Str. 77, 50937 K\"{o}ln, Germany}

\author{H. Kierspel}
    \affiliation{II. Physikalisches Institut, Universit\"{a}t zu K\"{o}ln, Z\"{u}lpicher Str. 77, 50937 K\"{o}ln, Germany}

\author{R. Sutarto}
\altaffiliation[Present address: ]{Canadian Light Source, University of Saskatchewan, Saskatoon, Saskatchewan S7N 0X4, Canada}
    \affiliation{II. Physikalisches Institut, Universit\"{a}t zu K\"{o}ln, Z\"{u}lpicher Str. 77, 50937 K\"{o}ln, Germany}

\author{L.H. Tjeng}
\email{Hao.Tjeng@cpfs.mpg.de}
    \affiliation{II. Physikalisches Institut, Universit\"{a}t zu K\"{o}ln, Z\"{u}lpicher Str. 77, 50937 K\"{o}ln, Germany}
\affiliation{Max Planck Institute for Chemical Physics of Solids, 01187 Dresden, Germany}

\author {A. Damascelli}
\email {damascelli@physics.ubc.ca}
\affiliation{Department of Physics and Astronomy, University of British Columbia, Vancouver, British Columbia V6T 1Z1, Canada}
\affiliation{Quantum Matter Institute, University of British Columbia, Vancouver, British Columbia V6T 1Z4, Canada}
\date{\today}

\newcommand{\degree}{\ensuremath{^\circ}}

\begin{abstract}
We have investigated a method for substituting oxygen with
nitrogen in EuO thin films, which is based on molecular beam
epitaxy distillation with $\ce{NO}$ gas as the oxidizer.
By varying the $\ce{NO}$ gas pressure, we produce crystalline,
epitaxial \ce{EuO_{1-x}N_x} films with
good control over the films' nitrogen concentration. In-situ x-ray
photoemission spectroscopy reveals that nitrogen substitution is
connected to the formation $\mathrm{Eu^{3+} \, 4\mathit{f}^6}$ and a
corresponding decrease in the number of $\mathrm{Eu^{2+} \,
4\mathit{f}^7}$, indicating that nitrogen is being incorporated in its $3-$
oxidation state. While small amounts of $\mathrm{Eu^{3+}}$ in
over-oxidized $\mathrm{Eu_{1-\delta}O}$ thin films lead to a
drastic suppression of the ferromagnetism, the formation of
$\mathrm{Eu^{3+}}$ in \ce{EuO_{1-x}N_x}
still allows the ferromagnetic phase to exist with an unaffected $\mathrm{T_c}$,
thus providing an ideal model system to study the interplay
between the magnetic $\mathit{f}^7$ ($\mathrm{J=7/2}$) and the non-magnetic
$\mathit{f}^6$ ($\mathrm{J=0}$) states close to the Fermi level.
\end{abstract}

\maketitle

The key to producing novel spintronic devices is to find magnetic
materials that can be combined with conventional semiconductors.
One class of materials being considered is that of the dilute
magnetic semiconductors (DMS). In this case, standard
semiconductors are doped with magnetic impurities, leading to
spin-dependent electron transport phenomena which stem from the
conduction-electron-mediated ferromagnetic
coupling\citep{Dietl-2000, Pearton-2004}. Another promising class
of materials is that of the ferromagnetic semiconductors (FMS)
from the rare earth pnictide and chalcogenide families. Despite
currently having lower Curie temperatures ($\mathrm{T_c}$) than the
DMSs, the rare earth compounds are interesting because of their
spectacular properties due to the interplay between the
large magnetic moments arising from the atomic-like f-orbitals and
the electrons in the wide conduction bands. Arguably the
archetypical FMS is EuO, with its $\mathrm{J=7/2}$ moment from the
$\mathrm{Eu^{2+} \mathit{f}^7}$ ions, a $\mathrm{T_c}$ of $\mathrm{69\, K}$,
with a spin-split conduction band which allows for spin tunneling with up to $100\%$ spin polarization
\citep{Steeneken-2002,Schmehl-2007,Panguluri-2008}. It exhibits a metal-insulator transition as a function of applied magnetic field and temperature with a resistivity jump of 6 and 8 orders of magnitude, respectively\citep{Oliver-1972, Shapira-1973,
Shapira-1973-2}; this is even higher than what is observed in the
colossal magneto-resistance manganites \citep{Ramirez-1997,
Imada-1998}. Furthermore, EuO was grown successfully on Si, GaN
and GaAs \citep{Lettieri-2003, Schmehl-2007,Swartz-2010, Caspers2011a}, and
could thus be readily incorporated in conventional semiconductor
technology. Similarly GdN, which is isoelectronic and
isostructural to EuO, exhibits many of the same interesting
properties with a $\mathrm{T_c}$ of about 60 K\citep{Li-1997,
Duan-2007}.

In the search for new compounds in the area of FMSs, and based on the
promising properties of EuO and GdN, another system that has
attracted some attention is EuN. At first glance this might appear
surprising, due to the non-magnetic, $\mathrm{J=0}$ character of
its $\mathrm{Eu^{3+} \mathit{f}^6}$ ions. However, recent band structure
calculations seem to indicate that the $\mathrm{J=0}$ state can be
spin polarized, giving rise to ferromagnetism with an unoccupied \textit{f} band located close to -- or even right at -- the chemical
potential \citep{Horne-2004, Aerts-2004, Johannes-2005}. This
would lead to the realization of half-metallic ferromagnetic
behavior. Ruck \textit{et al.} have investigated this possibility
in EuN films grown by molecular beam epitaxy (MBE). However, by
studying the magnetic properties by X-ray magnetic circular
dichroism, they found no evidence for a ferromagnetic state
\citep{Ruck-2011}.

In this work, in search of half-metallic ferromagnetic behavior associated with nominally non-magnetic $\mathrm{Eu^{3+}}$
ions, we follow a different approach. Rather than pure EuN, we
grow thin films of EuO$_{1-x}$N$_x$ aiming at embedding the
$\mathrm{Eu^{3+} \mathit{f}^6}$ ions induced by nitrogen substitution in
the well-defined ferromagnetic lattice of EuO. This is in accordance with the
above-mentioned rationale that the magnetization provided by the
EuO lattice on non-magnetic  $\mathrm{Eu^{3+} \,\mathit{f}^6}$ might lead
to the formation of a spin-polarized conduction band, whose
filling is directly controlled by the nitrogen concentration.
As a result of the narrow bandwidth of the \textit{f}-derived band, these
carriers would have by a large effective mass.

Since  $\mathrm{EuO_{1-x}N_x}$ has not been synthesized before, we must first determine if such a system can be grown. It is
widely known that it is difficult to grow pure EuO thin films. EuO
is extremely unstable in air, and even in UHV environment -- if it
is grown with too much supply of oxygen -- it will form
$\mathrm{Eu_2O_3}$ and/or $\mathrm{Eu_3O_4}$ phases. On the other
hand, if it is grown with too little supply of oxygen, Eu metal
clusters may form\citep{Altendorf-2011a}. Both of these situations
will deteriorate the extraordinary properties of EuO. These
problems can be overcome to successfully grow high-quality EuO
thin films by the MBE distillation method, which involves
evaporating europium metal onto a hot substrate in a low pressure
of
oxygen\citep{Steeneken-2002,Steeneken-thesis,Ulbricht-2008,Sutarto-2009}.
The low oxygen pressure prevents the formation of
$\mathrm{Eu_2O_3}$ and/or $\mathrm{Eu_3O_4}$, while any unreacted
metal is re-evaporated from the hot substrate surface thus maintaining the well-proven europium distillation growth technique \cite{Sutarto-2009, Altendorf-2011b, Caspers2011b}. This is the
approach we have chosen to grow our
\ce{EuO_{1-x}N_x} films; however, instead
of pure oxygen, we used $\ce{NO}$ gas, and during the course
of the work we also discovered that the oxygen-nitrogen substitution can be tuned.

\begin{figure}[t!] \centering
\includegraphics[width=3.4in]{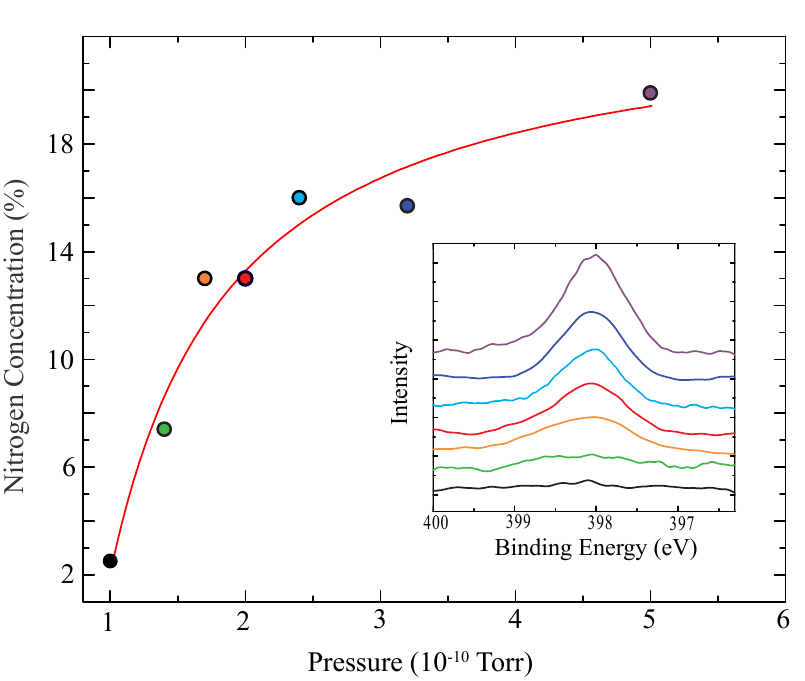}
\caption{Increase in nitrogen concentration in the
\ce{EuO_{1-x}N_x} films as a function of
$\ce{NO}$ partial pressure. These concentrations were measured
using the ratio of the background-subtracted nitrogen and oxygen
\textit{1s} core level areas in XPS, corrected for their
corresponding photo-ionization cross-sections. The inset shows the
increase in the nitrogen \textit{1s} core level peak with
increasing $\ce{NO}$ pressure as measured by XPS. The red line
is a guide to the eye, based on a fit to the Langmuir adsorption equation, see Ref. \onlinecite{Langmuir-1916}.}
\label{fig:NConcwPres}
\end{figure}

The choice of $\ce{NO}$ gas as both the oxidizer and as a
means of substituting nitrogen in EuO was inspired by earlier
$\mathrm{NO_2}$-assisted epitaxial growth of $\mathrm{Fe_3O_4}$,
$\mathrm{Fe_{1-\delta}O}$, and
$\mathrm{CrO}$\citep{Voogt-PRB-2001,Rogojanu-Thesis}. There,
$\mathrm{NO_2}$ gas was used because it is a very efficient
oxidizer; as a side effect it was found that, in addition to the
desired amount of oxygen, nitrogen was also being incorporated
into the films. Since the nitrogen concentration in the films was
decreasing upon increasing $\mathrm{NO_2}$ pressure, it was
hypothesized that the probability of nitrogen substitution was
higher when there was insufficient oxygen to form a stoichiometric
material. Since the conditions in MBE distillation are always
oxygen deficient by design, this technique can be used as a
general approach to incorporate nitrogen into oxide films. For the
case of EuO, since $\mathrm{NO_2}$ is far too aggressive an
oxidizer, we chose $\ce{NO}$ gas instead, following the work
on nitrogen substituted SrO by Elfimov \textit{et
al.}\citep{Elfimov-2007}. In that work, by keeping the rate of metal
evaporation constant, and changing the background pressure of
$\ce{NO}$ gas in the UHV growth chamber, it was shown that the amount of
nitrogen taken up by the film can be tuned.

The \ce{EuO_{1-x}N_x} samples were grown
on yttria-stabilized zirconia (YSZ) substrates, whose
$\mathrm{5.142 \AA}$ lattice constant is very close to one of bulk
EuO, $\mathrm{5.144 \AA}$. These substrates, purchased from
SurfaceNet GmbH., were annealed in the growth chamber for two hours
at $\mathrm{600 ^oC}$ in $\mathrm{1\times10^{-6}}$ Torr of oxygen
(the chamber base pressure is in the $10^{-10}$ Torr range). This
procedure removes surface contaminants, re-oxygenates the
substrate, and gives defects on the surface enough mobility to
aggregate into step edges, producing an atomically flat surface.
After annealing, the substrate temperature was set to $\mathrm{450
^oC}$. Europium metal was evaporated from a Knudsen cell at a rate of 8.2 \AA\, per minute; the rate was measured with a quartz crystal monitor. The chamber was
backfilled with $\ce{NO}$ gas through a precision leak valve
and the NO partial pressure was measured with a MKS Instruments
residual gas analyzer. While Eu evaporation rate and substrate
temperature were kept constant for all growths, the amount of
$\ce{NO}$ gas used to oxidize and dope the films was adjusted
for each growth. The range of gas pressures was between $\mathrm{1\times10^{-10}}$ and $\mathrm{5\times10^{-10}}$ Torr. The choice of substrate temperature, evaporation
rate, and $\ce{NO}$ pressure range determine if the conditions
are favourable for MBE distillation; starting
parameters were chosen based on earlier work \citep{Sutarto-2009,
SutartoGD-2009}.

\begin{figure}[b!] \centering
\includegraphics[width=3.4in]{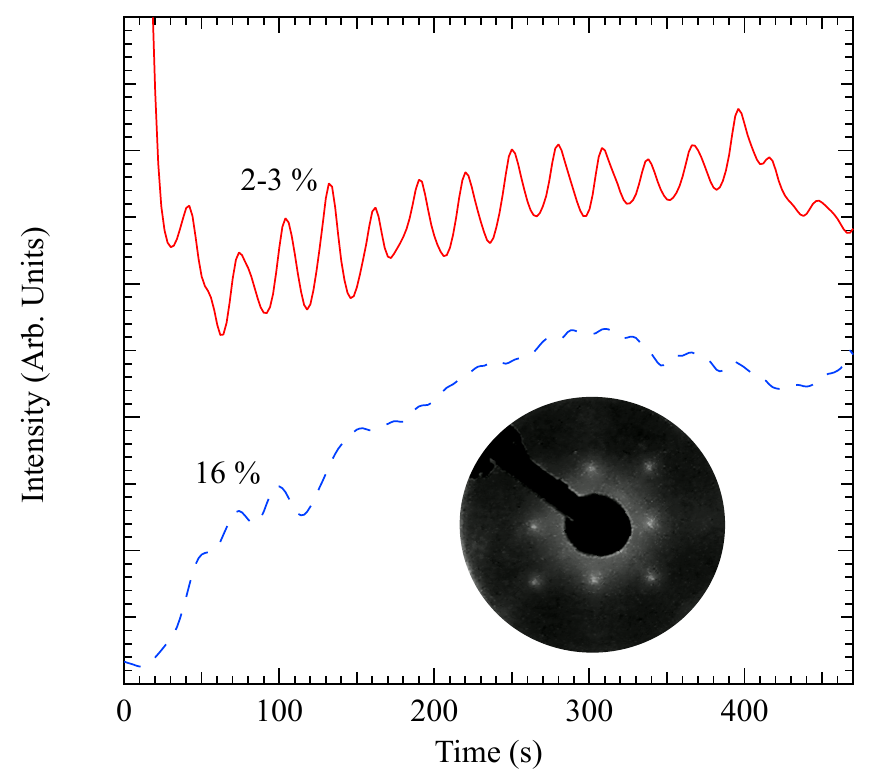}
\caption{RHEED spot intensity oscillations for $\mathrm{2-3 \%}$
and $\mathrm{16 \%} $ nitrogen concentration, respectively. The inset shows a LEED diffraction pattern for a $\mathrm{13 \%}$ N substituted sample. The
initial three oscillations in the $\mathrm{16 \%}$ nitrogen
substitution case are consistent with a layer-by-layer growth
process enabled by oxygen being donated to the film by the YSZ
substrate. The $\mathrm{2-3 \%}$ substitution film exhibits in
addition RHEED oscillations that continue well past the thickness
where oxygen donation from the substrate could have an effect,
indicating that these prolonged oscillations are due to sustained
layer-by-layer growth of \ce{EuO_{1-x}N_x}. The LEED diffraction pattern indicates that even after 100 minutes of growth, the film remains crystalline.} \label{RHEED}
\end{figure}

Fig.\,\ref{fig:NConcwPres} demonstrates how the amount of nitrogen
incorporated into the \ce{EuO_{1-x}N_x}
films varies by changing the background pressure of the
$\ce{NO}$ gas.  The
concentration of substituted nitrogen was estimated from the
photo-ionization cross-section\cite{ CrossSectionTheory} corrected ratio of nitrogen and oxygen \textit{1s}
core level peaks measured by x-ray photoemission spectroscopy
(XPS)\cite{Rogojanu-Thesis}. The XPS
measurements were performed in situ with monochromatized
$\mathrm{Al}\, K\alpha$ radiation and a VSW 150 electron analyzer.
As shown in the inset of Fig.\,\ref{fig:NConcwPres}, the nitrogen
\textit{1s} peak grows with the NO pressure (the spectra were
normalized to the oxygen \textit{1s} peak area, not shown).

We note that the non-linear increase of nitrogen concentration into the \ce{EuO_{1-x}N_x} films with increasing NO pressure is very different from the previously reported linear decrease with increasing gas pressure\citep{Voogt-PRB-2001,Rogojanu-Thesis}, observed in the $\mathrm{NO_2}$-assisted growth of $\mathrm{Fe_3O_4}$, $\mathrm{Fe_{1-\delta}O}$, and $\mathrm{CrO}$. In these latter cases, the metal-to-$\mathrm{NO_2}$ flux ratio was approximately $1$, setting the growth far outside of the MBE distillation regime used here. One could then envision that the $\mathrm{NO_2}$ oxidizes first the available Fe or Cr, leaving behind an equal number of $\ce{NO}$ molecules that can react with the remaining metal; since the number of these remaining metal sites is inversely proportional to the initial $\mathrm{NO_2}$ pressure, one obtains the observed linear decrease in nitrogen concentration with increasing $\mathrm{NO_2}$ pressure. For the present case of NO-assisted growth of EuO, we can conclude that the increase of the N-to-O ratio with increasing NO pressure stems specifically from the MBE distillation conditions, although further research will be needed for the accurate quantitative modeling of the adsorption kinetics under these conditions.

\begin{figure}[t!] \centering
\includegraphics[width=3.4in]{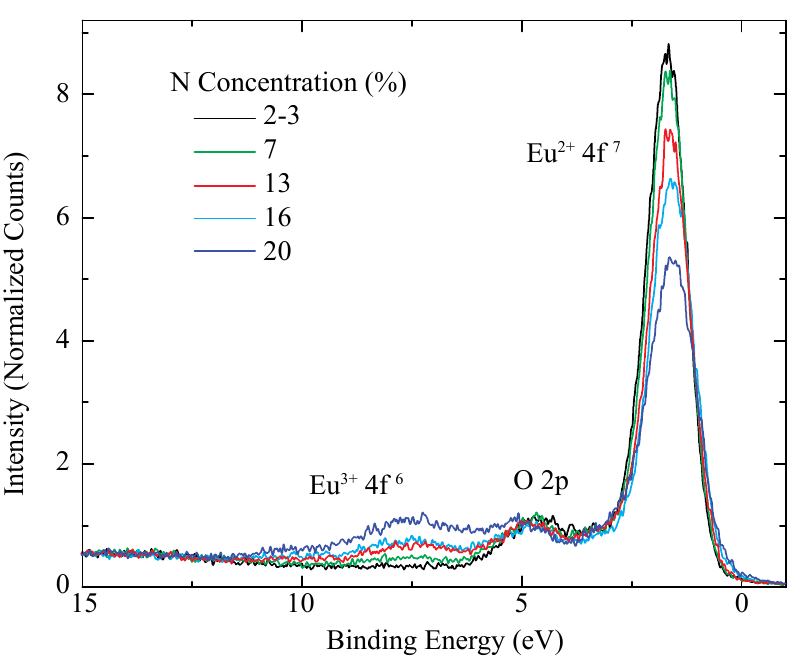}
\caption{Valence band XPS spectra of
\ce{EuO_{1-x}N_x} as a function of
nitrogen concentration. The increase of spectral intensity
in the $\mathrm{Eu^{3+}}$ peak, and the corresponding decrease
in $\mathrm{Eu^{2+}}$ spectral weight, suggest that nitrogen
is being incorporated in its $\mathrm{3^-}$ oxidation state.
The XPS spectra were normalized to the total number of counts.}
\label{fig:EuONVB}
\end{figure}

In addition to providing a method for introducing nitrogen into
EuO films, MBE distillation with $\ce{NO}$ gas also produces
films with excellent crystalline quality. All of the films
exhibited low-energy electron diffraction (LEED) and reflection
high-energy electron diffraction (RHEED) patterns, with
well-defined diffraction spots after growth. The films also
exhibited layer-by-layer growth under certain conditions, as
indicated by the presence of typical RHEED oscillations (see
Fig.\,\ref{RHEED} for representative RHEED data from
$\mathrm{2-3\%}$  and $\mathrm{16\%}$ nitrogen substituted
\ce{EuO_{1-x}N_x}, and crystalline LEED data from a $\mathrm{13\%}$ substituted sample). The $\mathrm{16\%}$
substitution level does exhibit three RHEED oscillations at
the beginning of growth, but they quickly disappear. These initial RHEED oscillations are seen in all the films, regardless of the substitution level;
however, the lower pressure growths exhibit RHEED oscillations
that continue for several 10's of monolayers. The difference
between high and low pressure regimes is most likely due to a too
high concentration of defects in the heavily nitrogen-substituted
films; these defects act as nucleation sites, initiating
three-dimensional island growth and destroying the
two-dimensional, layer-by-layer growth. Initially, however, the
growth is primarily controlled by oxygen being donated to the film
by the YSZ substrate, rather than by the $\ce{NO}$ gas
\citep{Sutarto-2009}: this allows the observation of
layer-by-layer growth independent of the $\ce{NO}$ gas
pressure. RHEED oscillations that continue beyond 4-5 monolayers
cannot be due to the substrate donating oxygen, because the film
is at that point too thick for oxygen from the substrate to
diffuse to the surface \citep{Sutarto-2009}. Therefore, these
additional oscillations must originate from
\ce{EuO_{1-x}N_x} growing in a truly
layer-by-layer mode, as in the case of the $\mathrm{2\!-\!3\%}$
substitution level shown in Fig.\,\ref{RHEED}.

\begin{figure}[t!] \centering
\includegraphics[width=3.4in]{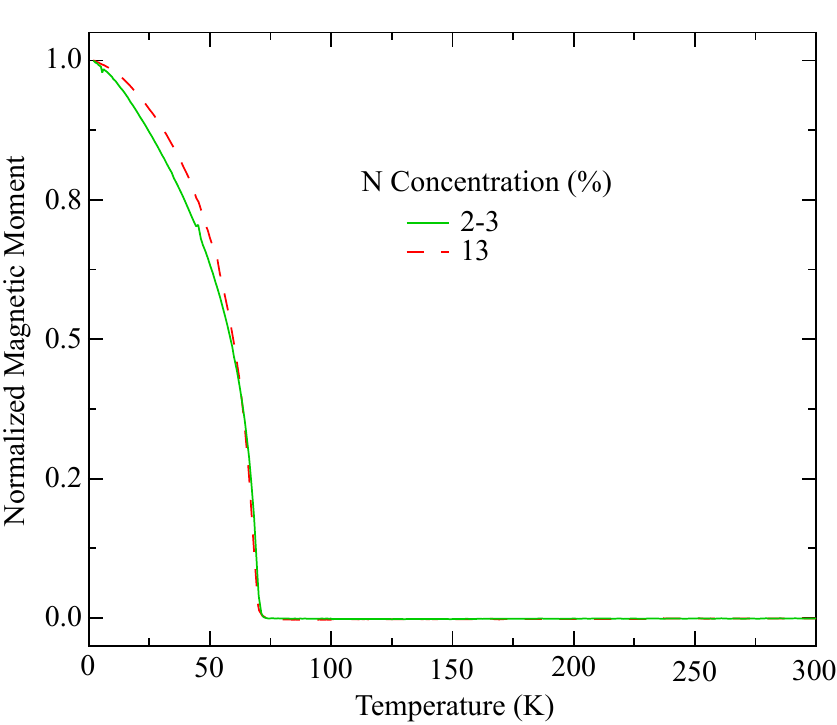}
\caption{Magnetization curves for two different
\ce{EuO_{1-x}N_x} samples at 10 Gauss. The transition
temperature remains unchanged from the EuO value of 69-70K, even
at high levels of nitrogen substitution.}
\label{fig:EuONMag}
\end{figure}

The results discussed above demonstrate that MBE distillation with
$\ce{NO}$ gas is a suitable approach for producing high
crystallinity, epitaxial \ce{EuO_{1-x}N_x}
films, with tunable nitrogen-oxygen substitution. To establish the
oxidation state of the substituted nitrogen, and provide a more
detailed characterization of the films' electronic structure, we
performed XPS valence band measurements in situ
(Fig.\,\ref{fig:EuONVB}). The spectra evolution indicates that as
the nitrogen concentration increases so does the amount of
$\mathrm{Eu^{3+}}$ spectral weight, while the $\mathrm{Eu^{2+}}$
intensity decreases. We also note that the oxygen 2p intensity does not increase with increasing NO-pressure, establishing that the increase of $\mathrm{Eu^{3+}}$ is due to the substitution of
$\mathrm{O^{2-}}$ with $\mathrm{N^{3-}}$, thus with nitrogen being incorporated in its 3- oxidation state.
Whereas these XPS results match those from over-oxidized
$\mathrm{Eu_{1-\delta}O}$ films, the electronic/magnetic
properties of  \ce{EuO_{1-x}N_x} are
remarkably different. In particular, at variance with the behavior
observed for $\mathrm{Eu_{1-\delta}O}$ where small amounts of
$\mathrm{Eu^{3+}}$ lead to a drastic suppression of the
ferromagnetism \citep{Sutarto-2009}, over-oxidation in
\ce{EuO_{1-x}N_x}still produces a lineshape of the magnetization curve which is Brillouin-like, suggesting an appreciable preservation of the ferromagnetic phase. Important is that the $\mathrm{T_c}$ of about 69 K, as in pure EuO, is observed over a wide range of nitrogen substitution. This is
demonstrated by magnetization measurements performed ex situ with
a Quantum Designs MPMS-XL7 SQUID magnetometer in a 10 Gauss field (after capping the
samples with a thick aluminum layer to protect them from further
oxidation when removed from the MBE system), and here shown in
Fig.\,\ref{fig:EuONMag}.

In conclusion, by substituting nitrogen for oxygen in EuO, we have
made a $\mathrm{Eu^{2+}}$/$\mathrm{Eu^{3+}}$ system
that remains ferromagnetic despite the inclusion of
$\mathrm{Eu^{3+} \, 4\mathit{f}^6}$ sites, something not possible in the
more extensively studied $\mathrm{Eu_{1-\delta}O}$. In addition,
\ce{EuO_{1-x}N_x} is also the ideal system
for the specific purpose of studying the hopping between the $\mathit{f}^7$
($\mathrm{J=7/2}$) and $\mathit{f}^6$ ($\mathrm{J=0}$) levels located in
proximity of the chemical potential. In this respect,
\ce{EuO_{1-x}N_x} is also better than
Eu$_x$Gd$_{1-x}$N -- in which such $f^6/f^7$ mixing is also achieved -- since in the latter case the $4\mathit{f}^7$ levels of
Eu and Gd are split in energy by several eV's
\cite{SutartoGD-2009}, preventing an efficient hopping within the
$\mathit{f}$ band. More generally, the MBE NO-assisted distillation
technique described here provides a means to tune the oxygen-nitrogen substitution in other binary oxides.

We thank I.S. Elfimov and G.A. Sawatzky for
discussions and suggestions, and L. Hamdan and T. Koethe for
technical assistance. We acknowledge support from the Max Plank - UBC Centre for Quantum Materials. The work at UBC was supported by the Killam,
Sloan, CRC, NSERC's Steacie Fellowship Programs (A.D.), NSERC,
CFI, CIFAR Quantum Materials, and BCSI. The research activities in
Cologne were supported by the Deutsche Forschungsgemeinschaft
through SFB 608.

%

\end{document}